%% file: main.tex
\documentclass[11pt]{article}

\newif\ifcomments

\commentstrue

\input{header}

\begin{document}

\title{A Note on the Hardness of Partially Dynamic Max Flow}


\author{
  Thatchaphol Saranurak\\
  University of Michigan \\
  \texttt{thsa@umich.edu}
  \and
  Benyu Wang\\
  University of Michigan\\
  \texttt{benyuw@umich.edu}
}

\date{}
\maketitle

\begin{abstract}
We study the fine-grained complexity of partially dynamic
max flow in capacitated graphs. Assuming the combinatorial
4-clique hypothesis, combinatorial algorithms require
$\Omega(n^{2-o(1)})$ amortized update time. Thus, recomputing
max flow after each update is optimal up to subpolynomial
factors for combinatorial algorithms in dense graphs.
The previous conditional lower bound was
$\Omega(n^{1-o(1)})$ amortized update time under SETH or the OMv
hypothesis and applies to all algorithms
\cite{dahlgaard2016hardness}.
    
\end{abstract}

\input{intro}

\input{prelim}

\input{lower}

\input{reduction}

\bibliographystyle{alpha}
\bibliography{ref}

\end{document}

%% file: header.tex
\usepackage{fullpage}
\usepackage[utf8]{inputenc}
\usepackage{tikz}
\usetikzlibrary{shapes.geometric, arrows}
\usepackage{graphicx}
\usepackage{amsmath,amsthm,amssymb}
\usepackage{algorithm}
\usepackage{algpseudocode}
\usepackage{multirow}
\usepackage{makecell}
\usepackage{algpseudocode}

\usepackage{thm-restate,color,xspace}
\usepackage{comment}
\usepackage{thmtools}
\usepackage{xcolor}
\usepackage{nameref}
\usepackage{array}
\usepackage{svg}

\definecolor{ForestGreen}{rgb}{0.1333,0.5451,0.1333}
\definecolor{DarkRed}{rgb}{0.65,0,0}
\definecolor{Red}{rgb}{1,0,0}
\usepackage[linktocpage=true,
pagebackref=true,colorlinks,
linkcolor=DarkRed,citecolor=ForestGreen,
bookmarks,bookmarksopen,bookmarksnumbered]
{hyperref}
\usepackage{cleveref}

\usepackage{xcolor}
\usepackage{nameref}

\usepackage{cleveref}

\makeatother

\usepackage{babel}
\usepackage{hyperref}

\declaretheorem[numberwithin=section]{theorem}
\declaretheorem[numberlike=theorem]{lemma}

\declaretheorem[numberlike=theorem,style=definition]{remark}

\newcommand{\flow}{\text{\normalfont flow}}

\newcommand{\ignore}[1]{}

\usepackage{todonotes}

\ifcomments
\def\thatchaphol#1{\marginpar{$\leftarrow$\fbox{TS}}\footnote{$\Rightarrow$~{\sf\textcolor{purple}{#1 --Thatchaphol}}}}
\def\benyu#1{\marginpar{$\leftarrow$\fbox{BW}}\footnote{$\Rightarrow$~{\sf\textcolor{red}{#1 --Benyu}}}}
\def\gary#1{\marginpar{$\leftarrow$\fbox{GH}}\footnote{$\Rightarrow$~{\sf\textcolor{blue}{#1 --Gary}}}}

\else
\newcommand{\benyu}[1]{}
\newcommand{\gary}[1]{}
\newcommand{\thatchaphol}[1]{}
\fi 

\usepackage[utf8]{inputenc} 
\DeclareUnicodeCharacter{1D405}{$\mathbf{F}$} 

\DeclareUnicodeCharacter{1D400}{\ensuremath{\mathbf{A}}}
\DeclareUnicodeCharacter{1D401}{\ensuremath{\mathbf{B}}}
\DeclareUnicodeCharacter{1D402}{\ensuremath{\mathbf{C}}}
\DeclareUnicodeCharacter{1D403}{\ensuremath{\mathbf{D}}}
\DeclareUnicodeCharacter{1D404}{\ensuremath{\mathbf{E}}}
\DeclareUnicodeCharacter{1D405}{\ensuremath{\mathbf{F}}}
\DeclareUnicodeCharacter{1D406}{\ensuremath{\mathbf{G}}}
\DeclareUnicodeCharacter{1D407}{\ensuremath{\mathbf{H}}}
\DeclareUnicodeCharacter{1D408}{\ensuremath{\mathbf{I}}}
\DeclareUnicodeCharacter{1D409}{\ensuremath{\mathbf{J}}}
\DeclareUnicodeCharacter{1D40A}{\ensuremath{\mathbf{K}}}
\DeclareUnicodeCharacter{1D40B}{\ensuremath{\mathbf{L}}}
\DeclareUnicodeCharacter{1D40C}{\ensuremath{\mathbf{M}}}
\DeclareUnicodeCharacter{1D40D}{\ensuremath{\mathbf{N}}}
\DeclareUnicodeCharacter{1D40E}{\ensuremath{\mathbf{O}}}
\DeclareUnicodeCharacter{1D40F}{\ensuremath{\mathbf{P}}}
\DeclareUnicodeCharacter{1D410}{\ensuremath{\mathbf{Q}}}
\DeclareUnicodeCharacter{1D411}{\ensuremath{\mathbf{R}}}
\DeclareUnicodeCharacter{1D412}{\ensuremath{\mathbf{S}}}
\DeclareUnicodeCharacter{1D413}{\ensuremath{\mathbf{T}}}
\DeclareUnicodeCharacter{1D414}{\ensuremath{\mathbf{U}}}
\DeclareUnicodeCharacter{1D415}{\ensuremath{\mathbf{V}}}
\DeclareUnicodeCharacter{1D416}{\ensuremath{\mathbf{W}}}
\DeclareUnicodeCharacter{1D417}{\ensuremath{\mathbf{X}}}
\DeclareUnicodeCharacter{1D418}{\ensuremath{\mathbf{Y}}}
\DeclareUnicodeCharacter{1D419}{\ensuremath{\mathbf{Z}}}

\DeclareUnicodeCharacter{1D41A}{\ensuremath{\mathbf{a}}}
\DeclareUnicodeCharacter{1D41B}{\ensuremath{\mathbf{b}}}
\DeclareUnicodeCharacter{1D41C}{\ensuremath{\mathbf{c}}}
\DeclareUnicodeCharacter{1D41D}{\ensuremath{\mathbf{d}}}
\DeclareUnicodeCharacter{1D41E}{\ensuremath{\mathbf{e}}}
\DeclareUnicodeCharacter{1D41F}{\ensuremath{\mathbf{f}}}
\DeclareUnicodeCharacter{1D420}{\ensuremath{\mathbf{g}}}
\DeclareUnicodeCharacter{1D421}{\ensuremath{\mathbf{h}}}
\DeclareUnicodeCharacter{1D422}{\ensuremath{\mathbf{i}}}
\DeclareUnicodeCharacter{1D423}{\ensuremath{\mathbf{j}}}
\DeclareUnicodeCharacter{1D424}{\ensuremath{\mathbf{k}}}
\DeclareUnicodeCharacter{1D425}{\ensuremath{\mathbf{l}}}
\DeclareUnicodeCharacter{1D426}{\ensuremath{\mathbf{m}}}
\DeclareUnicodeCharacter{1D427}{\ensuremath{\mathbf{n}}}
\DeclareUnicodeCharacter{1D428}{\ensuremath{\mathbf{o}}}
\DeclareUnicodeCharacter{1D429}{\ensuremath{\mathbf{p}}}
\DeclareUnicodeCharacter{1D42A}{\ensuremath{\mathbf{q}}}
\DeclareUnicodeCharacter{1D42B}{\ensuremath{\mathbf{r}}}
\DeclareUnicodeCharacter{1D42C}{\ensuremath{\mathbf{s}}}
\DeclareUnicodeCharacter{1D42D}{\ensuremath{\mathbf{t}}}
\DeclareUnicodeCharacter{1D42E}{\ensuremath{\mathbf{u}}}
\DeclareUnicodeCharacter{1D42F}{\ensuremath{\mathbf{v}}}
\DeclareUnicodeCharacter{1D430}{\ensuremath{\mathbf{w}}}
\DeclareUnicodeCharacter{1D431}{\ensuremath{\mathbf{x}}}
\DeclareUnicodeCharacter{1D432}{\ensuremath{\mathbf{y}}}
\DeclareUnicodeCharacter{1D433}{\ensuremath{\mathbf{z}}}

\DeclareUnicodeCharacter{1D434}{\ensuremath{\mathit{A}}}
\DeclareUnicodeCharacter{1D435}{\ensuremath{\mathit{B}}}
\DeclareUnicodeCharacter{1D436}{\ensuremath{\mathit{C}}}
\DeclareUnicodeCharacter{1D437}{\ensuremath{\mathit{D}}}
\DeclareUnicodeCharacter{1D438}{\ensuremath{\mathit{E}}}
\DeclareUnicodeCharacter{1D439}{\ensuremath{\mathit{F}}}
\DeclareUnicodeCharacter{1D43A}{\ensuremath{\mathit{G}}}
\DeclareUnicodeCharacter{1D43B}{\ensuremath{\mathit{H}}}
\DeclareUnicodeCharacter{1D43C}{\ensuremath{\mathit{I}}}
\DeclareUnicodeCharacter{1D43D}{\ensuremath{\mathit{J}}}
\DeclareUnicodeCharacter{1D43E}{\ensuremath{\mathit{K}}}
\DeclareUnicodeCharacter{1D43F}{\ensuremath{\mathit{L}}}
\DeclareUnicodeCharacter{1D440}{\ensuremath{\mathit{M}}}
\DeclareUnicodeCharacter{1D441}{\ensuremath{\mathit{N}}}
\DeclareUnicodeCharacter{1D442}{\ensuremath{\mathit{O}}}
\DeclareUnicodeCharacter{1D443}{\ensuremath{\mathit{P}}}
\DeclareUnicodeCharacter{1D444}{\ensuremath{\mathit{Q}}}
\DeclareUnicodeCharacter{1D445}{\ensuremath{\mathit{R}}}
\DeclareUnicodeCharacter{1D446}{\ensuremath{\mathit{S}}}
\DeclareUnicodeCharacter{1D447}{\ensuremath{\mathit{T}}}
\DeclareUnicodeCharacter{1D448}{\ensuremath{\mathit{U}}}
\DeclareUnicodeCharacter{1D449}{\ensuremath{\mathit{V}}}
\DeclareUnicodeCharacter{1D44A}{\ensuremath{\mathit{W}}}
\DeclareUnicodeCharacter{1D44B}{\ensuremath{\mathit{X}}}
\DeclareUnicodeCharacter{1D44C}{\ensuremath{\mathit{Y}}}
\DeclareUnicodeCharacter{1D44D}{\ensuremath{\mathit{Z}}}

\DeclareUnicodeCharacter{1D44E}{\ensuremath{\mathit{a}}}
\DeclareUnicodeCharacter{1D44F}{\ensuremath{\mathit{b}}}
\DeclareUnicodeCharacter{1D450}{\ensuremath{\mathit{c}}}
\DeclareUnicodeCharacter{1D451}{\ensuremath{\mathit{d}}}
\DeclareUnicodeCharacter{1D452}{\ensuremath{\mathit{e}}}
\DeclareUnicodeCharacter{1D453}{\ensuremath{\mathit{f}}}
\DeclareUnicodeCharacter{1D454}{\ensuremath{\mathit{g}}}
\DeclareUnicodeCharacter{210E}{\ensuremath{\mathit{h}}}
\DeclareUnicodeCharacter{1D456}{\ensuremath{\mathit{i}}}
\DeclareUnicodeCharacter{1D457}{\ensuremath{\mathit{j}}}
\DeclareUnicodeCharacter{1D458}{\ensuremath{\mathit{k}}}
\DeclareUnicodeCharacter{1D459}{\ensuremath{\mathit{l}}}
\DeclareUnicodeCharacter{1D45A}{\ensuremath{\mathit{m}}}
\DeclareUnicodeCharacter{1D45B}{\ensuremath{\mathit{n}}}
\DeclareUnicodeCharacter{1D45C}{\ensuremath{\mathit{o}}}
\DeclareUnicodeCharacter{1D45D}{\ensuremath{\mathit{p}}}
\DeclareUnicodeCharacter{1D45E}{\ensuremath{\mathit{q}}}
\DeclareUnicodeCharacter{1D45F}{\ensuremath{\mathit{r}}}
\DeclareUnicodeCharacter{1D460}{\ensuremath{\mathit{s}}}
\DeclareUnicodeCharacter{1D461}{\ensuremath{\mathit{t}}}
\DeclareUnicodeCharacter{1D462}{\ensuremath{\mathit{u}}}
\DeclareUnicodeCharacter{1D463}{\ensuremath{\mathit{v}}}
\DeclareUnicodeCharacter{1D464}{\ensuremath{\mathit{w}}}
\DeclareUnicodeCharacter{1D465}{\ensuremath{\mathit{x}}}
\DeclareUnicodeCharacter{1D466}{\ensuremath{\mathit{y}}}
\DeclareUnicodeCharacter{1D467}{\ensuremath{\mathit{z}}}

\DeclareUnicodeCharacter{1D7CE}{\ensuremath{\mathbf{0}}}
\DeclareUnicodeCharacter{1D7CF}{\ensuremath{\mathbf{1}}}
\DeclareUnicodeCharacter{1D7D0}{\ensuremath{\mathbf{2}}}
\DeclareUnicodeCharacter{1D7D1}{\ensuremath{\mathbf{3}}}
\DeclareUnicodeCharacter{1D7D2}{\ensuremath{\mathbf{4}}}
\DeclareUnicodeCharacter{1D7D3}{\ensuremath{\mathbf{5}}}
\DeclareUnicodeCharacter{1D7D4}{\ensuremath{\mathbf{6}}}
\DeclareUnicodeCharacter{1D7D5}{\ensuremath{\mathbf{7}}}
\DeclareUnicodeCharacter{1D7D6}{\ensuremath{\mathbf{8}}}
\DeclareUnicodeCharacter{1D7D7}{\ensuremath{\mathbf{9}}}
\DeclareUnicodeCharacter{2265}{\ensuremath{\geq}}     
\DeclareUnicodeCharacter{2212}{\ensuremath{-}}        
\DeclareUnicodeCharacter{2032}{\ensuremath{^\prime}}  
\DeclareUnicodeCharacter{2113}{\ensuremath{\ell}}     
\DeclareUnicodeCharacter{22EF}{\ensuremath{\cdots}}   
\DeclareUnicodeCharacter{2033}{\ensuremath{^{\prime\prime}}}   
\DeclareUnicodeCharacter{2034}{\ensuremath{^{\prime\prime\prime}}} 

%% file: intro.tex
\section{Introduction}

Maximum flow is one of the most fundamental problems in algorithmic graph theory. Recently, Chen et al.~\cite{chen2025maximum} gave an almost-linear-time maximum-flow algorithm in the static setting. We study the \emph{partially dynamic} max-flow problem, where the goal is to maintain the maximum-flow value as the graph undergoes only edge insertions or only edge deletions. Throughout, $n$ denotes the number of vertices and $m$ the maximum number of edges in the graph.

A trivial approach is to recompute the flow from scratch in $m^{1+o(1)}$ time after each update, giving an $n^{2+o(1)}$ bound for dense graphs. Dahlgaard~\cite{dahlgaard2016hardness} proved an $\Omega(n^{1−o(1)})$ conditional lower bound on update time under SETH~\cite{impagliazzo2001complexity} or the OMv hypothesis~\cite{henzinger2015unifying}. This still leaves open whether an algorithm can beat recomputation with update time $O(n^{2-\epsilon})$ for some constant $\epsilon>0$.

\paragraph{Our Results.}
We rule out this update time for combinatorial algorithms under the \emph{combinatorial 4-clique hypothesis} (see, e.g.,~\cite{bringmann2017dichotomy}), which states that every combinatorial algorithm for finding a 4-clique in an $n$-vertex graph requires $\Omega(n^{4-o(1)})$ time\footnote{Here, \emph{combinatorial algorithms} generally refers to algorithms that do not use fast
matrix multiplication. The best known combinatorial algorithm runs in $O(n^4/2^{\Omega(\log^{1/7} n)})$ time~\cite{abboud2024new}. In contrast, Dahlgaard's lower bound~\cite{dahlgaard2016hardness} applies to \emph{all} algorithms under different hypotheses.}.

\begin{theorem}
    \label{thm:main}
    Assuming the combinatorial 4-clique hypothesis, every combinatorial algorithm that solves the partially dynamic max-flow problem in directed or undirected capacitated graphs with $O(n^2)$ edges requires $\Omega(n^{4-o(1)})$ total update time over a sequence of $O(n^2)$ updates.
\end{theorem}

The proof uses a simple reduction from \emph{all-pairs max flow} in \emph{directed graphs}: computing the maximum-flow value for every ordered pair of distinct vertices $(u,v)$. Under the combinatorial 4-clique hypothesis, this problem requires $\Omega(n^{4-o(1)})$ time~\cite{abboud2018faster}.

For \emph{undirected graphs}, all-pairs max flow can be solved in almost-linear time $m^{1+o(1)}$~\cite{gomory1961multi,abboud2023all}. For partially dynamic max flow, however, we show that the directed and undirected variants are equally hard.

\paragraph{Related Work.} For $(1-\epsilon)$-approximate max flow, partially dynamic algorithms achieve $m^{1+o(1)}$ total update time~\cite{chen2024almost,van2024almost}. Thus, the \emph{exact} and \emph{approximate} versions of partially dynamic max flow have sharply different complexities.

%% file: prelim.tex
\section{Preliminaries}

All graphs are capacitated and may be directed or undirected. Parallel edges are allowed. We describe only the \emph{incremental} max-flow problem; the lower bound for the \emph{decremental} variant follows by reversing the update sequence.

Let $G=(V,E)$ undergo edge insertions, and fix a source-sink pair $(s,t)$. For simplicity, assume that $G$ is initially edgeless. We write $G+e$ for the graph obtained by inserting an edge $e$ into $G$. The \emph{incremental max-flow} problem asks for the maximum-flow value from $s$ to $t$, denoted by $\flow_G(s,t)$, after each update.

We prove hardness by reducing from \emph{all-pairs max flow} in directed graphs. Given a directed graph $H=(V,E)$, this problem asks for $\flow_H(u,v)$ for every ordered pair of distinct vertices $u,v\in V$. We use the following lower bound.

\begin{theorem}[\cite{abboud2018faster}]
    \label{thm:APMF}
    Assuming the combinatorial 4-clique hypothesis, every combinatorial algorithm that solves the all-pairs max-flow problem in directed graphs requires $\Omega(n^{4-o(1)})$ time.
\end{theorem}

%% file: lower.tex
\section{Hardness of Incremental Max-Flow in Directed Graphs}

We first prove \Cref{thm:main} for incremental max-flow in directed graphs. Given a directed all-pairs max-flow instance $H=(V,E)$, we show how to compute all values $\flow_H(u,v)$ using a single incremental max-flow instance $G$.

Let $W$ be an upper bound on all values $\flow_H(u,v)$. We initialize an edgeless directed graph $G$ with vertex set $V \cup \{s,t\}$, where $s$ and $t$ are a new source and sink. Following \cite{dahlgaard2016hardness}, we use the following update sequence:

\begin{itemize}
    \item Insert every edge in $H$ with its original capacity.
    \item For each ordered pair of distinct vertices $(u,v)$, perform the following two steps:
    \begin{itemize}
        \item Insert edges $(s,u)$ and $(v,t)$ with capacity $W$, then read the maintained value $\flow_G(s,t)$.
        \item Insert edges $(u,t)$ and $(s,v)$ with capacity $W$.
    \end{itemize}
\end{itemize}

We show how the values read during these $O(n^2)$ updates determine $\flow_H(u,v)$ for every vertex pair $(u,v)$. The key observation is the following lemma.

\begin{lemma}
    \label{lem:twoedges}
    For any graph $G$ with source-sink pair $(s,t)$ and any vertex $u \in G$, inserting edges $(s,u)$ and $(u,t)$ with capacity $W$ gives $\flow_{G+(s,u)+(u,t)}(s,t) = \flow_{G}(s,t) + W$.
\end{lemma}

\input{figures/section3_add_edges}

\begin{proof}
    See \Cref{fig:paired-terminal-edges}. In $G+(s,u)+(u,t)$, first send $W$ units of flow along the path $s \to u \to t$. The residual graph has the same max-flow value as $G$: the reverse edges $(u,s)$ and $(t,u)$ cannot appear in any acyclic $(s,t)$ flow. The statement follows.
\end{proof}

\input{figures/section3_shortcuts}

\begin{proof}[Proof of \Cref{thm:main} in directed graphs]
    See \Cref{fig:shortcut-completed-pairs}. For any vertex pair $(u,v)$ in $H$, the $(s,t)$ max-flow value in $H+(s,u)+(v,t)$ equals $\flow_H(u,v)$, since both added edges have capacity $W \geq \flow_H(u,v)$. Consider the moment just after $(s,u)$ and $(v,t)$ are inserted, before the second step for $(u,v)$. Suppose $p$ vertex pairs precede $(u,v)$ in the sequence. For each such pair $(u',v')$, we have already inserted all four edges $(s,u')$, $(u',t)$, $(s,v')$ and $(v',t)$. Repeatedly applying \Cref{lem:twoedges} to the edge pairs $(s,u'),(u',t)$ and $(s,v'),(v',t)$ gives
    \[\flow_{G}(s,t) = \flow_{H+(s,u)+(v,t)} (s,t) + 2pW = \flow_{H}(u,v)+2pW.\]
    Thus, subtracting $2pW$ from the maintained value $\flow_{G}(s,t)$ gives $\flow_{H}(u,v)$. Doing this for every pair solves directed all-pairs max-flow. By \Cref{thm:APMF}, this yields the $\Omega(n^{4-o(1)})$ lower bound under the combinatorial 4-clique hypothesis.
\end{proof}

%% file: figures/section3_add_edges.tex
\begin{figure}[h]
  \centering
  \begin{tikzpicture}[
    font=\small,
    vertex/.style={circle,draw=black,fill=white,minimum size=6mm,
                   inner sep=0pt,font=\small},
    internal/.style={circle,draw=gray!65,fill=white,minimum size=3mm,
                     inner sep=0pt},
    original/.style={->,>=latex,draw=gray!65,line width=.65pt},
    inserted/.style={->,>=latex,draw=blue!65!black,line width=1.15pt},
    reverse/.style={->,>=latex,draw=red!65!black,dashed,line width=1pt},
    capacity/.style={fill=white,inner sep=1.5pt,font=\small}
  ]
    \foreach \offset/\panel in {0/left,7/right} {
      \begin{scope}[xshift=\offset cm]
        \draw[rounded corners=9pt,draw=gray!35,fill=gray!3]
          (-2.95,-1.5) rectangle (2.95,1.5);
        \node[text=gray!65!black,anchor=north west]
          at (-2.8,1.42) {$G$};
        \node[vertex] (\panel-s) at (-2.35,0) {$s$};
        \node[vertex] (\panel-u) at (0,0) {$u$};
        \node[vertex] (\panel-t) at (2.35,0) {$t$};
        \node[internal] (\panel-a) at (-.85,.95) {};
        \node[internal] (\panel-b) at (.85,.95) {};
        \node[internal] (\panel-c) at (-.85,-.95) {};
        \node[internal] (\panel-d) at (.85,-.95) {};
        \draw[original] (\panel-s) -- (\panel-a);
        \draw[original] (\panel-a) -- (\panel-b);
        \draw[original] (\panel-b) -- (\panel-t);
        \draw[original] (\panel-s) -- (\panel-c);
        \draw[original] (\panel-c) -- (\panel-d);
        \draw[original] (\panel-d) -- (\panel-t);
        \draw[original] (\panel-a) -- (\panel-u);
        \draw[original] (\panel-u) -- (\panel-d);
      \end{scope}
    }

    \node[font=\small\bfseries] at (0,2.02)
      {(a) Insert $(s,u)$ and $(u,t)$};
    \draw[inserted] (left-s) -- node[capacity,above=2pt] {$W$} (left-u);
    \draw[inserted] (left-u) -- node[capacity,above=2pt] {$W$} (left-t);
    \node[align=center,text width=5.9cm] at (0,-1.94)
      {Send $W$ units along the new path $s\to u\to t$.};

    \node[font=\small\bfseries] at (7,2.02)
      {(b) Residual graph after this flow};
    \draw[reverse] (right-u) --
      node[capacity,above=2pt] {$W$} (right-s);
    \draw[reverse] (right-t) --
      node[capacity,above=2pt] {$W$} (right-u);
    \node[align=center,text width=5.9cm] at (7,-1.94)
      {The new reverse arcs cannot occur in any acyclic $s$--$t$ flow.};
  \end{tikzpicture}
  \caption{Inserting a pair of edges at $u$.}
  \label{fig:paired-terminal-edges}
\end{figure}

%% file: figures/section3_shortcuts.tex
\begin{figure}[h]
  \centering
  \begin{tikzpicture}[
    font=\small,
    vertex/.style={circle,draw=black,fill=white,minimum size=6mm,
                   inner sep=0pt,font=\small},
    internal/.style={circle,draw=gray!65,fill=white,minimum size=3mm,
                     inner sep=0pt},
    original/.style={->,>=latex,draw=gray!65,line width=.7pt},
    probe/.style={->,>=latex,draw=blue!65!black,line width=1.1pt},
    shortcut/.style={->,>=latex,draw=orange!80!black,line width=1.1pt},
    capacity/.style={fill=white,inner sep=1.5pt,font=\small}
  ]
    \foreach \offset/\panel in {0/left,7/right} {
      \begin{scope}[xshift=\offset cm]
        \draw[rounded corners=9pt,draw=gray!35,fill=gray!3]
          (-1.65,-.98) rectangle (1.65,.98);
        \node[anchor=north west,text=gray!65!black]
          at (-1.52,.91) {$H$};
        \node[vertex] (\panel-s) at (-2.65,0) {$s$};
        \node[vertex] (\panel-u) at (-.95,0) {$u$};
        \node[vertex] (\panel-v) at (.95,0) {$v$};
        \node[vertex] (\panel-t) at (2.65,0) {$t$};
        \node[internal] (\panel-a) at (-.4,.55) {};
        \node[internal] (\panel-b) at (.4,.55) {};
        \node[internal] (\panel-c) at (-.4,-.55) {};
        \node[internal] (\panel-d) at (.4,-.55) {};
        \draw[original] (\panel-u) -- (\panel-a);
        \draw[original] (\panel-a) -- (\panel-b);
        \draw[original] (\panel-b) -- (\panel-v);
        \draw[original] (\panel-u) -- (\panel-c);
        \draw[original] (\panel-c) -- (\panel-d);
        \draw[original] (\panel-d) -- (\panel-v);
        \draw[probe] (\panel-s) --
          node[capacity,above=2pt] {$W$} (\panel-u);
        \draw[probe] (\panel-v) --
          node[capacity,above=2pt] {$W$} (\panel-t);
      \end{scope}
    }

    \node[font=\small\bfseries,align=center] at (0,2.6)
      {(a) Add $(s,u)$ and $(v,t)$};

    \begin{scope}[xshift=7cm]
      \node[font=\small\bfseries,align=center] at (0,2.6)
        {(b) Shortcut $u$ and $v$};
      \draw[shortcut] (right-u)
        .. controls (-.55,1.85) and (2.2,1.85) ..
        node[capacity,above=3pt] {$W$} (right-t);
      \draw[shortcut] (right-s)
        .. controls (-2.2,-1.85) and (.55,-1.85) ..
        node[capacity,below=3pt] {$W$} (right-v);
    \end{scope}
  \end{tikzpicture}
  \caption{Checking $\flow_H(u,v)$ and shortcutting.}
  \label{fig:shortcut-completed-pairs}
\end{figure}
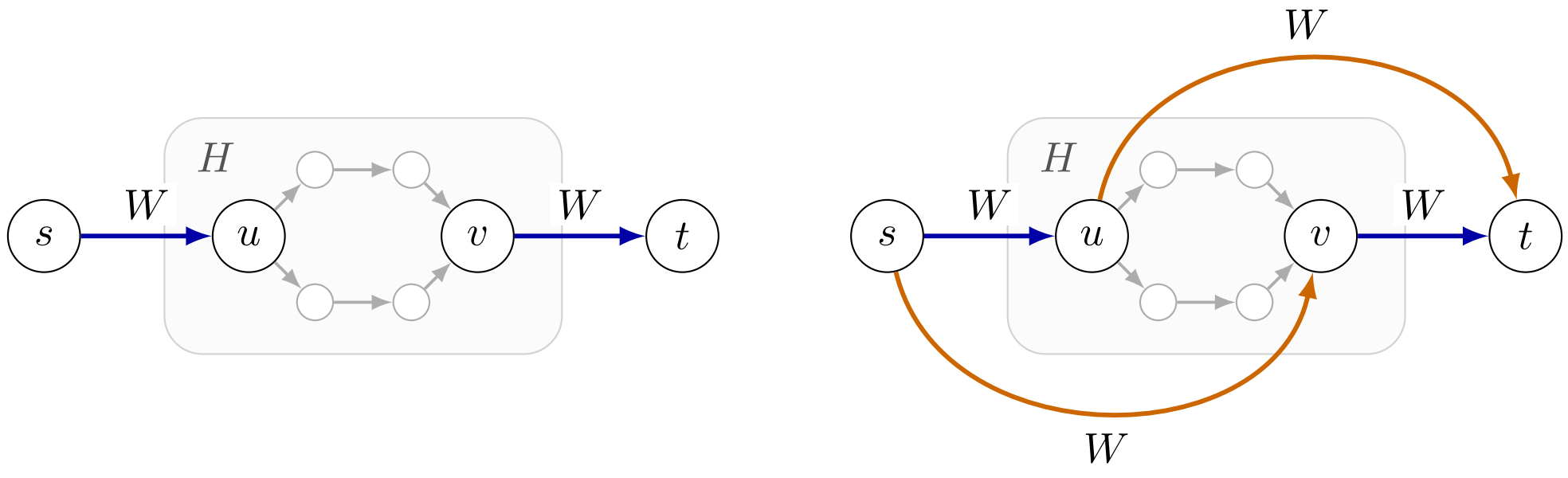

%% file: reduction.tex
\section{Reduction to Undirected Graphs}

We complete the proof of \Cref{thm:main} using a standard reduction from directed to undirected incremental max flow (see, e.g., Theorem 3.6.1 in \cite{mkadry2011graphs}). 

We first give a static reduction for $(s,t)$ max flow. Let $D$ be a directed graph with terminals $s$ and $t$. We construct an undirected graph $G$ as follows: for each directed edge $(u,v) \in D$ with capacity $w(u,v)$, we add three undirected edges $sv$, $vu$, and $ut$, each with capacity $w(u,v)$\footnote{If $\{u,v\} \cap \{s,t\} \neq \emptyset$, this construction may create loops, which do not affect the statement.}. The maximum-flow values satisfy the following relation:

\begin{lemma}
    \label{lem:static}
    $\flow_G(s,t) = 2\flow_D(s,t) + \sum_{(u,v) \in D} w(u,v)$.
\end{lemma}

\input{figures/section4_three_edges}

\begin{proof}
    See \Cref{fig:three-edge-reduction}. For each edge $(u,v) \in D$, we first send $w(u,v)$ units of flow along the corresponding edges $sv$, $vu$, and $ut$ in $G$. This flow has value $\sum_{(u,v) \in D} w(u,v)$. Let $R$ be its directed residual graph. Then $\flow_G(s,t) = \flow_R(s,t) + \sum_{(u,v) \in D} w(u,v)$.
    
    In $R$, each directed edge $(u,v) \in D$ gives three directed edges $(v,s)$, $(u,v)$, and $(t,u)$, each with capacity $2w(u,v)$. We can delete the first and third edges contributed by each gadget without changing the $(s,t)$ max-flow value, since they do not appear in any acyclic $(s,t)$ flow. The remaining graph has one edge $(u,v)$ of capacity $2w(u,v)$ for each edge $(u,v) \in D$. Therefore, $\flow_R(s,t) = 2 \flow_D(s,t)$. Combining the two equations proves the lemma.
\end{proof}

\begin{proof}[Proof of \Cref{thm:main} in undirected graphs]
    Let $D$ be an initially edgeless directed graph under edge insertions. We simulate $D$ with an undirected graph $G$: for each insertion of a directed edge $(u,v)$ into $D$, we insert three edges $(s,v)$, $(v,u)$, and $(u,t)$ into $G$, each with the same capacity as $(u,v)$. An incremental algorithm for undirected graphs maintains $\flow_G(s,t)$. By \Cref{lem:static}, we can recover $\flow_D(s,t)$ after each insertion by also maintaining the sum of the capacities in $D$. Thus, the $\Omega(n^{4-o(1)})$ conditional lower bound also holds for undirected graphs.
\end{proof}

\begin{remark}
    The construction above uses $\Theta(n^2)$ insertions. For sparser graphs with $m=o(n^2)$, an $\Omega(m^{2-o(1)})$ lower bound follows by adding isolated vertices in the standard way\footnote{See, e.g., Corollary 3.10 in \cite{huang2023tight}.}. Indeed, we can perform the above construction with $m$ edges on $q = \Theta(\sqrt{m})$ vertices, leaving the other $n-q$ vertices isolated. The resulting lower bound is $\Omega(q^{4-o(1)}) = \Omega(m^{2-o(1)})$. Thus, the amortized insertion time is still $\Omega(m^{1-o(1)})$.
\end{remark}


\paragraph{Conclusion and Current Landscape.}
We settle the $m^2$ total update complexity of partially
dynamic max flow in \emph{capacitated} graphs for both undirected and directed setting, up to subpolynomial
factors, for combinatorial algorithms under the combinatorial
4-clique hypothesis.

For \emph{unit-capacity simple} graphs, the maximum-flow
value is at most $n-1$. Hence, existing IPM-based algorithms
give $mn^{1+o(1)}$ total update time
\cite{chen2024almost,van2024almost}, while a simple
augmenting-path algorithm gives $O(mn)$ total update time
in the incremental setting~\cite{gupta2018simple}.
On dense instances, these bounds match Dahlgaard's
$\Omega(mn^{1-o(1)})$ conditional lower bound~\cite{dahlgaard2016hardness}.

\paragraph{Open Problem: Unit-Capacity Multigraphs.}
The complexity remains unresolved for \emph{unit-capacity
multigraphs}, where the maximum-flow value can be $\Theta(m)$.
Does partially dynamic max flow in this setting admit an
$\Omega(m^{2-o(1)})$ conditional lower bound on total update
time, or an algorithm with $O(m^{2-\epsilon})$ total update
time for some constant $\epsilon>0$? This remains an open problem.

\paragraph{AI Disclosure.} 
All technical ideas are by the authors. 
The authors used GPT 6.1-Sol to prepare the figures in this paper and to polish the grammar.

%% file: figures/section4_three_edges.tex
\begin{figure}[h]
  \centering
  \begin{tikzpicture}[
    font=\small,
    vertex/.style={circle,draw=black,fill=white,minimum size=6mm,
                   inner sep=0pt,font=\small},
    directed/.style={->,>=latex,draw=blue!65!black,line width=1.05pt},
    undirected/.style={draw=black,line width=.85pt},
    initialflow/.style={->,>=latex,draw=orange!80!black,line width=.9pt},
    discarded/.style={->,>=latex,draw=gray!70,dashed,line width=.9pt},
    capacity/.style={fill=white,inner sep=1.3pt,font=\small}
  ]
    \begin{scope}
      \node[font=\small\bfseries,align=center] at (0,1.43)
        {(a) Directed graph $D$};
      \node[vertex] (du) at (-1,0) {$u$};
      \node[vertex] (dv) at (1,0) {$v$};
      \draw[directed] (du) -- node[capacity,above=2pt] {$w$} (dv);
      \node[align=center,text width=4.1cm] at (0,-1.1)
        {Capacity $w$.};
    \end{scope}


    \begin{scope}[xshift=5cm]
      \node[font=\small\bfseries,align=center] at (0,1.43)
        {(b) Undirected graph $G$};
      \node[vertex] (gs) at (-1.8,0) {$s$};
      \node[vertex] (gv) at (-.6,0) {$v$};
      \node[vertex] (gu) at (.6,0) {$u$};
      \node[vertex] (gt) at (1.8,0) {$t$};
      \draw[undirected] (gs) -- node[capacity,above=2pt] {$w$} (gv);
      \draw[undirected] (gv) -- node[capacity,above=2pt] {$w$} (gu);
      \draw[undirected] (gu) -- node[capacity,above=2pt] {$w$} (gt);
      \draw[initialflow] (-1.57,-.44) -- (-.83,-.44);
      \draw[initialflow] (-.37,-.44) -- (.37,-.44);
      \draw[initialflow] (.83,-.44) -- (1.57,-.44);
      \node[align=center,text width=4.4cm] at (0,-1.1)
        {Take $w$ units of flow\\along $s\to v\to u\to t$.};
    \end{scope}


    \begin{scope}[xshift=10cm]
      \node[font=\small\bfseries,align=center] at (0,1.43)
        {(c) Residual graph $R$};
      \node[vertex] (rs) at (-1.8,0) {$s$};
      \node[vertex] (rv) at (-.6,0) {$v$};
      \node[vertex] (ru) at (.6,0) {$u$};
      \node[vertex] (rt) at (1.8,0) {$t$};
      \draw[discarded] (rv) -- node[capacity,above=2pt] {$2w$} (rs);
      \draw[directed] (ru) -- node[capacity,above=2pt] {$2w$} (rv);
      \draw[discarded] (rt) -- node[capacity,above=2pt] {$2w$} (ru);
      \node[align=center,text width=4.4cm] at (0,-1.1)
        {Delete $(v,s)$ and $(t,u)$.\\Obtain $(u,v)$ (capacity $2w$).};
    \end{scope}
  \end{tikzpicture}
  \caption{The reduction from directed to undirected maximum flow.}
  \label{fig:three-edge-reduction}
\end{figure}
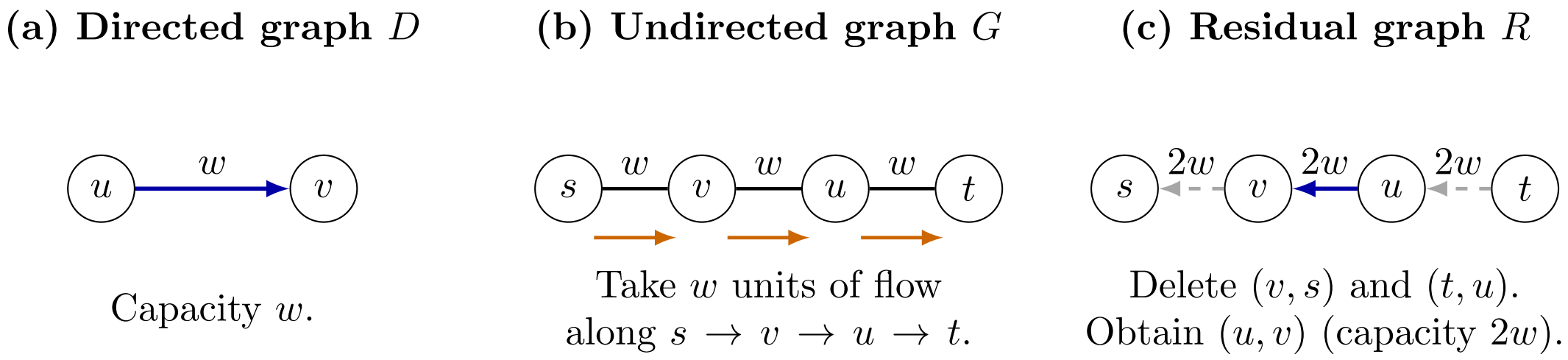